\documentclass[useAMS,usenatbib]{mn2e}
\usepackage{graphicx}
\usepackage{amsmath,amssymb}
\usepackage{color}
\usepackage{bm}

\newcommand\lsim{\mathrel{\rlap{\lower4pt\hbox{\hskip1pt$\sim$}}
        \raise1pt\hbox{$<$}}}
\newcommand\gsim{\mathrel{\rlap{\lower4pt\hbox{\hskip1pt$\sim$}}
        \raise1pt\hbox{$>$}}}

\title[A method to search for recoiling SMBHs in AGNs]{A statistical method to search for recoiling supermassive black holes in active galactic nuclei}
\author[Raffai, Haiman, \& Frei]{P. Raffai$^{1,2}$\thanks{E-mail: praffai@bolyai.elte.hu (PR);}, Z. Haiman$^{3}$, and Z. Frei$^{1,2}$ \\
	$^1$Institute of Physics, E\"otv\"os University, 1117 Budapest, Hungary \\
        $^2$MTA-ELTE EIRSA ``Lend\"ulet'' Astrophysics Research Group, 1117 Budapest, Hungary \\
        $^3$Department of Astronomy, Columbia University, New York, NY 10027, USA}

\begin{document}

\date{Released 2015 Xxxxx XX}

\pagerange{\pageref{firstpage}--\pageref{lastpage}} \pubyear{2015}

\maketitle

\label{firstpage}

\begin{abstract}
We propose an observational test for gravitationally recoiling
supermassive black holes (BHs) in active galactic nuclei, based on a
correlation between the velocities of BHs relative to their host
galaxies, $|\Delta v|$, and their obscuring dust column densities,
$\Sigma_\mathrm{dust}$ (both measured along the line of
sight). We use toy models for the distribution
of recoil velocities, BH trajectories, and the geometry of obscuring
dust tori in galactic centres, to simulate $2.5\times 10^5$ random
observations of recoiling quasars. BHs with recoil velocities
comparable to the escape velocity from the galactic centre remain
bound to the nucleus, and do not fully settle back to the centre of
the torus due to dynamical friction in a typical quasar lifetime.
We find that $|\Delta v|$ and $\Sigma_\mathrm{dust}$ for these BHs are
positively correlated. For obscured ($\Sigma_\mathrm{dust}>0$) and
for partially obscured ($0<\Sigma_\mathrm{dust}\lsim
2.3\ \mathrm{g/m^2}$) quasars with $|\Delta v|\geq 45\ \mathrm{km/s}$,
the sample correlation coefficient between $\log_{10}(|\Delta v|)$ and
$\Sigma_\mathrm{dust}$ is $r_{45} = 0.28\pm 0.02$ and $r_{45} =
0.13\pm 0.02$, respectively. Allowing for random $\pm 100\rm{km/s}$
errors in $|\Delta v|$ unrelated to the recoil dilutes the 
correlation for the partially obscured quasars to $r_{45} = 0.026\pm 0.004$ measured between $|\Delta v|$ and
$\Sigma_\mathrm{dust}$. 
A random sample of $\gsim 3,500$ obscured quasars with $|\Delta v|\geq
45\ \mathrm{km/s}$ would allow rejection of the no-correlation
hypothesis with $3\sigma$ significance $95$ per cent of the time. Finally,
we find that the fraction of obscured quasars, $\cal{F_{\rm
    obs}}$$\left (|\Delta v|\right )$, decreases with $|\Delta v|$
from $\cal{F_{\rm obs}}$$\left (<10\ \mathrm{km/s}\right )\gtrsim 0.8$
to $\cal{F_{\rm obs}}$$\left (>10^3\ \mathrm{km/s}\right )\lesssim
0.4$. This predicted trend can be compared to the observed fraction of
type II quasars, and can further test combinations of recoil,
trajectory, and dust torus models.
\end{abstract}


\section{Introduction}
\label{sec:Introduction}

Both theoretical models 
\citep{Begelmanetal1980,Volonterietal2003} and observations
\cite[e.g.][]{Comerfordetal2009} suggest that it is common for
supermassive black holes (SMBHs) in galactic centres to form binaries
that gradually lose their energy through radiating gravitational waves
\citep[e.g.][]{Haehnelt1994,Sesana2005}. Numerical simulations of
black hole (BH) binary mergers (see e.g. \citealt{Healyetal2014} and
references therein) suggest that the merger remnant SMBHs receive a
recoil velocity of typically several hundred (and in some spin and
mass-ratio configurations up to thousands) of km/s, due to highly
anisotropic gravitational-wave emission in the final merger
phase. Taking into account the spatial distribution of mass in the
galactic centre region, and corresponding dynamical friction, this
means that the SMBH can engage in a damped oscillating motion
\citep{MadauQuataert2004,KomossaMerritt2008,TanakaHaiman2009} with an
amplitude comparable to, or exceeding the $\cal{O}$(10-100 pc) size of
the optically thick dusty molecular torus (`dust torus') believed to
be surrounding galactic centres (for an overview, see
\citealt{Honig2008}). For a large recoil, the SMBH can escape from the
galactic centre and can remain wandering within the dark matter halo
\citep{Guedesetal2009,Guedesetal2011}. As accreting material can stay
gravitationally bound to the moving SMBH, and thus the SMBH can remain
active for $10^7-10^8$ yr after the merger event
(e.g. \citealt{Loeb2007}), kinematic and spatial signatures of the
recoil could be found in the spectra of quasars 
(QSOs; see e.g. \citealt{Bonningetal2007,Blechaetal2011,Komossa2012,Blechaetal2015}). These
signatures could confirm the gravitational recoil of merger remnant
SMBHs, and could probe the SMBH binary parameters, recoil trajectory
models, as well as the spatial distribution and composition of matter
in the galactic centre region.

In this paper, we propose that in the presence of dust tori obscuring
galactic nuclei, the gravitational recoil of active merger remnant
SMBHs should introduce a correlation between dust column mass
densities along the line of sight ($\Sigma_\mathrm{dust}$) and
magnitudes of line-of-sight peculiar velocities of SMBHs relative to
their host galaxies ($|\Delta v|$). Proxies to estimate both of these
quantities can be measured from observable features of QSO spectra
\citep{Bonningetal2007,Ledouxetal2015}. As pointed out earlier by
\citet{KomossaMerritt2008}, recoiling BHs can spend a significant
fraction of their time off-nucleus, possibly reducing the fraction of
fully obscured (`type II') QSOs; we also follow up on this
suggestion.

Using a selected combination of models of gravitational recoil, SMBH
trajectories, and obscuring dust tori, we demonstrate the feasibility
of detecting $\Sigma_\mathrm{dust}$--$|\Delta v|$ correlations by
simulating a set of $2.5\times 10^5$ random observations of recoiled
QSOs with a Monte Carlo method. We characterize the strength of
correlation between $\Sigma_\mathrm{dust}$ and $|\Delta v|$ for
obscured (i.e. $\Sigma_\mathrm{dust}>0$) QSOs in two different
$|\Delta v|$ intervals, and estimate the number of obscured QSOs that
could be used to reject the hypothesis of no correlation between
$\Sigma_\mathrm{dust}$ and $|\Delta v|$ with $3\sigma$
significance. As the strength of the correlation between
$\Sigma_\mathrm{dust}$ and $|\Delta v|$, as well as the underlying
$\Sigma_\mathrm{dust}\left (|\Delta v| \right )$ relation depends on
the presumed combination of models, observational studies on QSO
spectra could provide an opportunity for testing chosen combinations
of these models. We also calculate the fraction of QSOs obscured by
their dust tori, $\cal{F_{\rm obs}}$$\left (|\Delta v|\right )$,
which, compared to the observed fraction of type II (i.e. obscured)
QSOs could provide an independent test for a chosen combination of
recoil, SMBH trajectory, and dust tori models. We propose to use the
SDSS-DR10 Quasar Catalog \citep{Parisetal2014} to perform these tests
in the near future, and will report on this observational search in a
separate publication.

The paper is organized as follows. In \S~\ref{sec:Dynamics}, we
describe the gravitational recoil and SMBH trajectory models we used
in our Monte Carlo simulations. In \S~\ref{sec:Torus}, we describe our
implementation of a smooth dust torus model. We present our results in
\S~\ref{sec:Results} and discuss some issues that will be relevant for
an observational search for the proposed correlations in
\S~\ref{sec:Discussion}.  Finally, we offer our conclusions and
summarize the implications of this work in \S~\ref{sec:Conclusions}.

\section{BH Dynamics}
\label{sec:Dynamics}

Anisotropic emission of gravitational waves emitted by coalescing
SMBHs carry away linear momentum, resulting in a recoil of the merger
remnant SMBH in the opposite direction. Numerical simulations of this
process have been carried out for merging SMBHs with equal and unequal
masses, zero and non-zero spins aligned or counter-aligned with the
orbital angular momentum, and with spins pointing in random directions
with equal probability (see e.g. \citealt{Healyetal2014} and
references therein).

To keep our Monte Carlo simulation as general and realistic, but at
the same time, as computationally cheap as possible, following
\citet{TanakaHaiman2009}, we adopted the analytical formulae given in
equation (4) of \citet{Bakeretal2008} to construct the distribution of
recoil velocity magnitudes, $v_\mathrm{recoil}$ (note that
\citealt{Bakeretal2008} gives results very similar to more recent,
slightly modified formulae in \citealt{Loustoetal2012} and \citealt{Healyetal2014}).

The direction of the
recoil velocity depends on the orientation of the spin and orbital
angular momentum vectors. It is plausible that prior to merger, the
BHs accrete a significant amount of gas coherently from a circumbinary
accretion disc, whose inner regions are aligned with the binary's
orbital plane (e.g \cite{Ivanovetal99}). One then expects that the
spin angular momentum vectors at the time of the merger may be aligned
with the orbital angular momentum of the binary
(\citealt{Bogdanovicetal2007}; although there could still be
significant misalignment at merger for fast-spinning and unequal-mass
binaries; \citealt{Gerosaetal2015}). Thus, the kick direction may not
be random, and may lie preferentially in the plane of the
circumbinary disc (and also in the binary orbital plane). However, we
emphasize that here we are interested only in the kick direction
relative to the orientation of the larger scale nuclear torus. It is
much less clear whether the binary orbital plane and/or accretion disc
is aligned with the symmetry plane of this torus. This depends on the
transfer of angular momentum between large (100 pc) and small (sub-parsec)
scales, which is sensitive to turbulence, star formation, and feedback
processes \citep{Duboisetal2014}. Observationally, parsec-scale maser
discs (which could be taken as proxies for the accretion discs that
determine the kick direction) appear to be oriented randomly with
respect to the plane of their host galaxies \citep{KormendyHo2013}.
We therefore simply assume that the kick direction is random, i.e. not
aligned with the symmetry plane of the torus. Thus, we chose directions of recoil velocities randomly from a
uniform distribution covering a whole sphere. 

Using the fitting formulae of \citet{Bakeretal2008}, we calculated
$v_\mathrm{recoil}$ for a given pair of masses ($m_{1,2}$) and
dimensionless spin vectors of the merging SMBHs
($\vec{\alpha}_{1,2}\equiv c\vec{S}_{1,2}/Gm_{1,2}^2$, where
$\vec{S}_{1,2}$ are the spins of the SMBHs, $c$ is speed of light, and
$G$ is Newton's constant). Observations of active galactic nuclei
(AGNs) indicate that a significant number of SMBHs have
$|\vec{\alpha}_{1,2}|>0.9$, although a second population of SMBHs with
$0.4<|\vec{\alpha}_{1,2}|<0.8$ was found for $m_{1,2}>4\times 10^7
M_{\odot}$ SMBHs (see \citealt{Reynolds2013}). In our simulations, for simplicity, we first assigned each merging
SMBH to one of the two $|\vec{\alpha}_{1,2}|$ populations with equal
probability, and then drew a random $|\vec{\alpha}_{1,2}|$ value from
the corresponding (i.e. $|\vec{\alpha}_{1,2}|\in [0.9,1]$ or
$|\vec{\alpha}_{1,2}|\in [0.4,0.8]$) interval. The directions of both spin
vectors were chosen independently from a uniform distribution covering
a whole sphere. 

\begin{figure}
  \begin{center}
  \includegraphics[width=84mm]{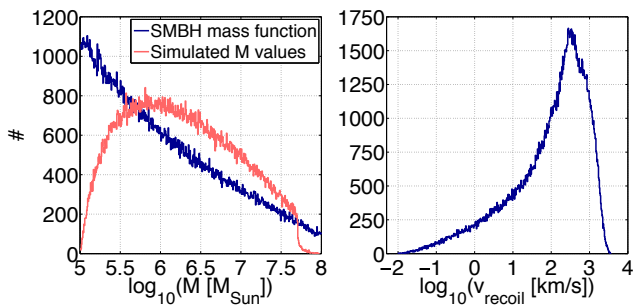}\\
   \caption{Left: histogram of $2.5\times 10^5$ simulated merger
     remnant SMBH masses $M=m_{1}+m_{2}$ (red curves), where the
     values of $2m_{1}$ and $2m_{2}$ were randomized from the SMBH
     mass function given in \citet{AllerRichstone2002} and represented
     by dark blue curves. As a result of the randomization process,
     $M$ values are under-represented at the lowest ($\sim 10^5
     M_{\odot}$) and at the highest ($\sim 10^8 M_{\odot}$) values,
     and overrepresented at values in between, compared to the SMBH
     mass function. Right: histogram of the corresponding recoil
     velocity magnitudes, $v_\mathrm{recoil}$, of the merger remnant
     SMBHs, obtained using the framework presented in
     \citet{Bakeretal2008}, and assuming that merging SMBHs have
     dimensionless spin magnitudes, $|\vec{\alpha}_{1,2}|$, from two
     distinct populations (i.e. $|\vec{\alpha}_{1,2}|\in [0.9,1]$ and
     $|\vec{\alpha}_{1,2}|\in [0.4,0.8]$; see \citealt{Reynolds2013})
     with equal probability, that $|\vec{\alpha}_{1,2}|$ values are
     uniformly distributed within each of the two
     $|\vec{\alpha}_{1,2}|$ intervals, and that $\vec{\alpha}_{1,2}$
     have uniformly distributed random directions.}
  \label{Fig:vrecoil}
 \end{center}
 \end{figure}

We randomized pairs of masses independently from the
SMBH mass function given in \citet{AllerRichstone2002}, downscaled
both mass values by a factor of 2, and used them as values for
$m_{1}$ and $m_{2}$. The fact that we chose the mass of the merger
remnant as $M=m_{1}+m_{2}$ means that in our Monte Carlo simulations
$M$ values are under-represented at the lowest ($\sim 10^5 M_{\odot}$)
and at the highest ($\sim 10^8 M_{\odot}$) values, and overrepresented
at values in between, compared to the SMBH mass function given in
\citet{AllerRichstone2002}. Fig.~\ref{Fig:vrecoil} shows the
histograms $M$ and $v_\mathrm{recoil}$ values we obtained by
randomizing $2.5\times 10^5$ pairs of merging SMBHs.

\begin{figure}
  \begin{center}
  \includegraphics[width=84mm]{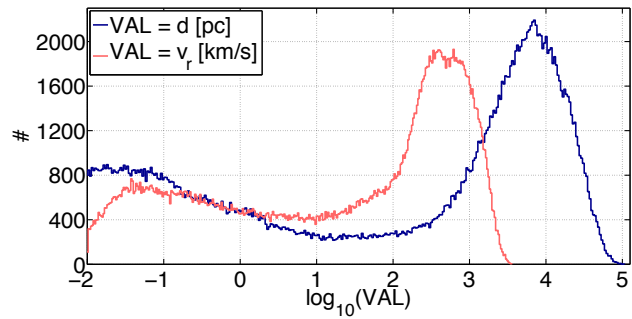}\\
   \caption{Histogram of distance $d$ (in pc; blue curve), and
     radial velocity magnitude, $v_\mathrm{r}$ (in km/s; red curve),
     of the $2.5\times 10^5$ SMBHs, both measured from the galactic
     centre at the randomly picked time of a mock observation. Both
     histograms show bimodality. The peaks at lower $d$ and
     $v_\mathrm{r}$ values
     ($\log_{10}(\mathrm{VAL})<-1$) correspond to
     SMBHs that underwent at least a half period of oscillation and
     thus suffered a strong orbital decay due to dynamical friction in
     the galactic centre. The peaks at high ($\log_{10}(\mathrm{VAL})>2$)
     $d$ and $v_\mathrm{r}$ values correspond to SMBHs whose initial recoil velocity was high enough
     to make them gravitationally unbound and escape from the galactic
     centre region. }
  \label{Fig:rv}
 \end{center}
 \end{figure}

Using the above set of $v_\mathrm{recoil}$'s with uniformly
distributed random directions, we simulated the resulting SMBH
trajectories based on the model presented in
\citet{MadauQuataert2004}.  The initial position of the SMBH and the
origin of our coordinate system were chosen to coincide with the
galactic centre. This model assumes that the SMBH is embedded in a
spherical stellar bulge, and is decelerated by dynamical friction.
The one-dimensional velocity dispersion of stars in the bulge were
calculated using the empirical $M$--$\sigma$ relation
\citep{Tremaineetal2002} as $\sigma_{1D} = \left( 1/\sqrt{3}
\right)\times \left( M/ 1.3\times 10^8 {\rm M_{\odot}} \right )^{1/4}
200\ \mathrm{km/s}$. According to observations by
\citet{BarthGreeneHo2005} and \citet{GreeneHo2006}, this relation
extends to active SMBHs with masses as low as $M\sim 10^5
M_{\odot}$. We simulated the trajectories with a $\Delta
t=10^3\ \mathrm{yr}$ time resolution up to a randomly picked final
time between $T\in [\Delta t, 3\times 10^4\Delta t]$. The maximum
duration of $T_\mathrm{max} \equiv 3\times 10^4\Delta t = 3\times
10^7\ \mathrm{yr}$ was chosen to be comparable with the maximum known
duration of QSO activity (see \citealt{Martini2004} for a review). The
simulation was terminated before $T$ if the orbit of the SMBH has
decayed and the SMBH settled back at the galactic centre (i.e. if the
radial distance, $d$, and radial velocity, $v_\mathrm{r}$, of the SMBH
converged to $d<10^{-2}\ \mathrm{pc}$ and
$v_\mathrm{r}<10^{-2}\ \mathrm{km/s}$, respectively). The position,
velocity, and $\Delta v$ of the SMBH in its final state were
calculated and stored, as if they were results of a QSO observation
made at a random time during the QSO activity after the
recoil. Histograms of $d$ and $v_\mathrm{r}$ values obtained for the
$2.5\times 10^5$ QSOs are shown in Fig.~\ref{Fig:rv}. Note that we
assumed that recoiled SMBHs remain active along their entire
trajectory, and also that they are active at the time the mock
observation is made.

 \citet{Guedesetal2009} followed the motions of merger remnant SMBHs
 with fixed masses in various configurations of dark matter haloes
 using numerical simulations. They concluded that due to the
 asymmetric and inhomogeneous mass distribution in the halo, the SMBH
 trajectories suffer large deviations from the
 \citet{MadauQuataert2004} trajectories. SMBHs might not even return
 to the galactic centre, after the SMBHs reach the first turning point
 in their oscillating motion. To check how much this could affect the
 results of our Monte Carlo simulations, we simulated an independent
 set of $10^5$ QSOs, and examined the fraction of SMBHs that reach the
 first turning point before the mock observation is made.  We found
 that for the entire QSO sample, this fraction is $\simeq 31$ per cent, while
 it is reduced to $\simeq 11$ and $\simeq 4$ per cent for QSOs with
 $|\Delta v|\geq 5\ \mathrm{km/s}$ and $|\Delta v|\geq
 45\ \mathrm{km/s}$, respectively.  
 The reason for the full population having a higher fraction is that
 once the SMBHs make their first U--turn, their orbits decay rapidly
 due to dynamical friction. As a result, many of these end up with
 $|\Delta v|< 5\ \mathrm{km/s}$ and $|\Delta v|< 45\ \mathrm{km/s}$ at
 the mock observation time (as shown in Fig.~\ref{Fig:rv}).  As we
 will show in \S~\ref{sec:Results} below, only QSOs with $|\Delta
 v|\geq 5\ \mathrm{km/s}$ show a $\Sigma_\mathrm{dust}$--$|\Delta v|$
 correlation. Furthermore, below we propose to study only the subset
 of QSOs with $|\Delta v|\geq 45\ \mathrm{km/s}$, due to the
 limitations imposed by inevitable random velocity errors.  We
 conclude that in this sub-sample of QSOs, the effects of asphericity
 and inhomogeneities will reduce the predicted correlations only by
 $\approx 4$ per cent, and we ignore this complication in the rest of this
 paper.
 
A similar deviation from a purely radial trajectory may arise on
much smaller scales, inside the spherical nuclear star cluster (stellar core)
enshrouding the galactic centre. As shown by \citet{GualandrisMerritt2008},
when the SMBH mass is not negligible compared to the mass
of the stellar core, the SMBH and the stellar core exhibit
oscillations about their common centre of mass. As a result,
the SMBH is not exposed to the largest dynamical friction near
the geometric centre of the core, and it may take up to
$\sim 10$ times longer for the oscillations to be damped. Here,
we neglect this effect, leaving its evaluation to future work. 
However, we note that it would likely result in a larger fraction of
SMBHs remaining displaced from the nucleus, and possibly a larger
fraction of QSOs exhibiting the correlations we propose.

\section{Torus model}
\label{sec:Torus}

According to the unified scheme of AGNs, galactic nuclei are obscured
by optically and geometrically thick dusty molecular tori, with the
amount of obscuration depending on the viewing angle
\citep{Antonucci1993}. Even though dust typically only constitutes
$\sim 1$ per cent of the mass of the tori (while the rest of the mass is in
gas, mostly in molecular hydrogen form), dust is the main component
responsible for the obscuration and reddening of QSOs, and thus, most
torus models focus on the dust distribution in the tori (i.e. on
`dust tori').  

Torus models developed to reproduce the observed spectral energy
distributions of AGNs can be divided into two categories: smooth versus
clumpy models (for an overview, see e.g. \citealt{Honig2008}).  We
implemented a hydrostatic model of smooth dust tori presented by
\citet{Schartmannetal2005}. This model has the advantages of
simplicity, using physically reasonable assumptions about torus
formation, and providing good fits to the mid-infrared spectral energy
distributions of AGNs with relatively few free parameters. The dust
torus is assumed to form from gas released through stellar winds and
ejection of planetary nebulae in the nuclear star cluster. Even though the
ejecta of individual stars should produce a cloudy structure of the
torus, no instruments so far has been able to resolve single clouds of
the dust distribution, and \citet{Schartmannetal2005} conclude that
these clouds should be small. For simplicity, they treat the torus as
a continuous medium characterized by the density distribution
$\rho_\mathrm{d}$ (see their equation 8). The effective potential created
by the central SMBH and the angular momentum distribution in the
stellar core makes $\rho_\mathrm{d}$ axisymmetric around the galactic
centre.

To cover the range of masses $M_{*}$ of the nuclear star clusters in
the examples in \citet{Schartmannetal2005} and the observations
presented in \citet{Leighetal2012}, we chose
$M_{*}=10^{5+\beta}M_{\odot}$, where $\beta$ was drawn from a uniform
distribution in the range $\beta\in [0,4]$. We set the mass of the
dust enclosed in the torus to $M_\mathrm{dust} = (5.79\times 10^5
M_{\odot})\times [M_{*}/(2\times 10^9 M_{\odot})]$ in order to
reproduce parameter values of the example torus given in table~1 in
\citet{Schartmannetal2005}.

The radius of the stellar core, $R_\mathrm{c}$, was chosen to be
\begin{equation}\label{eq:Rc}
 R_\mathrm{c} = 2\times 10^{(a\log_{10} [M_{*}/M_{\odot}] - b)}\ \mathrm{pc},
\end{equation}
where the dimensionless constants $a=0.3042$ and $b=1.2679$ were
obtained by fitting the half-light core radii,
$r_\mathrm{eff}=0.5R_\mathrm{c}$, for late- and early-type galaxies in
Figure~13 in \citet{GeorgievBoker2014}.
Note that we assumed a constant stellar mass-to-light ratio,
$M_{*}/L_\mathrm{V}$, where $L_\mathrm{V}$ is the total luminosity of
the nuclear star cluster in the $V$ band.

The radius $R_\mathrm{T}$ of the torus was chosen such that the sizes
of the torus and of the nuclear star cluster are comparable, with
$R_\mathrm{T}=5\ \mathrm{pc}$ for $M_{*}=2\times 10^9 M_{\odot}$ (see
table~1 in \citealt{Schartmannetal2005}):
\begin{equation}\label{eq:RT}
 R_\mathrm{T} = 3.56\times10^{-2}\ R_\mathrm{c} + 2.33\ \mathrm{pc} .
\end{equation}

For all other parameters in our torus simulations, we followed the
methods in \citet{Schartmannetal2005}. Specifically, we set the outer
radius of the torus $R_\mathrm{out}=3R_\mathrm{c}$, the exponent of the angular momentum distribution in the stellar core $\gamma=-0.5$, and the turbulent velocity of the clouds building up the torus $v_\mathrm{t}\approx \sigma_{*}$, where we used equation (9) in
\citet{Schartmannetal2005} to calculate the velocity dispersion of the stars in the nuclear star cluster, $\sigma_{*}$.

The density distribution of a torus in this model is fully determined
by two parameters: the mass of the nuclear star cluster, $M_{*}$, and
the mass of the central SMBH, $M$. Using the random values of $M_{*}$
and $M$, for each torus we calculated
$\hat{\rho}_\mathrm{d}(R,z)\equiv
\rho_\mathrm{d}(R,z)/\rho_\mathrm{d}^0$ (see Eq.~8 in
\citealt{Schartmannetal2005}) in a $R_\mathrm{out}\times
R_\mathrm{out}$ rectangular cross-section along the $R-z$ plane with a
linear resolution of $R_\mathrm{out}/200$, set
$\hat{\rho}_\mathrm{d}(R,z)=0$ wherever $\hat{\rho}_\mathrm{d}(R,z) <
\hat{\rho}_\mathrm{d}(R_\mathrm{out},0)$, and calculated
$\rho_\mathrm{d}^0$ such that the total mass of the torus with the
resulting $\rho_\mathrm{d}(R,z)$ equals $M_\mathrm{dust}$. In our
simulation, only the dust column mass density, $\Sigma_\mathrm{dust}$,
was recorded as the final output, which was calculated by numerically
integrating $\rho_\mathrm{d}(R,z)$ along the line of sight from each
SMBH position to the observer for a randomly oriented torus.

\begin{figure}
  \begin{center}
  \includegraphics[width=84mm]{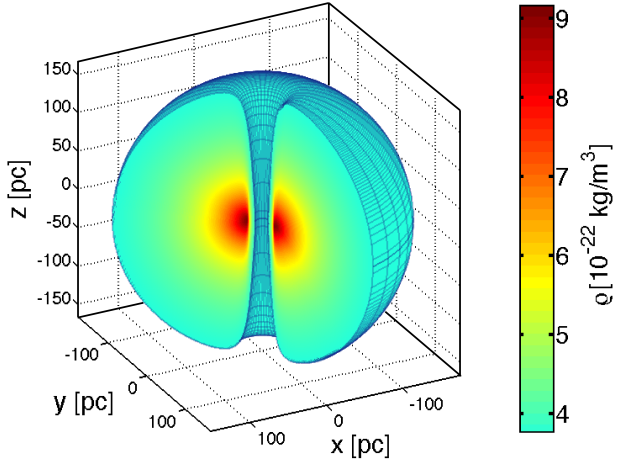}\\
   \caption{An example for the adopted geometry and interior density
     structure along a cross-section of a torus surrounding an AGN, 
     based on \citet{Schartmannetal2005}. The
     mass of the nuclear star cluster containing the torus and
     producing its dust content was chosen to be $M_{*} = 10^{9}
     M_{\odot}$, while the mass of the central SMBH was set to
     $M = 10^6 M_{\odot}$.}
  \label{Fig:Torus}
 \end{center}
 \end{figure}

\begin{figure}
  \begin{center}
  \includegraphics[width=84mm]{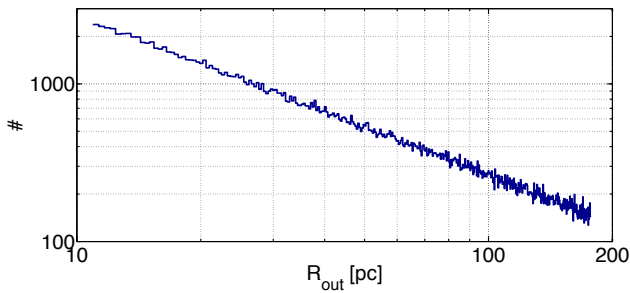}\\
   \caption{Histogram of outer radii, $R_\mathrm{out}$, of the
     $2\times 10^{5}$ simulated dust tori. $R_\mathrm{out}$ values
     follow a $\propto 1/R_\mathrm{out}$ distribution, ranging from
     $R_\mathrm{out,min} \simeq 10\ \mathrm{pc}$ to
     $R_\mathrm{out,max} \simeq 180\ \mathrm{pc}$.}
  \label{Fig:Rout}
 \end{center}
 \end{figure}

For illustration, Fig.~\ref{Fig:Torus} shows a visualization of the
geometry and interior density structure ($\rho_\mathrm{d}$) along a
cross-section of a torus with $M_{*} = 10^{9} M_{\odot}$ and $M = 10^6
M_{\odot}$.  Additionally, in Fig.~\ref{Fig:Rout}, we show the
histogram of the outer radii, $R_\mathrm{out}$, of the $2\times
10^{5}$ dust tori. This shows that the outer radii follow a $\propto
1/R_\mathrm{out}$ distribution, ranging from $R_\mathrm{out,min}
\simeq 10\ \mathrm{pc}$ to $R_\mathrm{out,max} \simeq
180\ \mathrm{pc}$.

\vspace{-\baselineskip}
\section{Results}
\label{sec:Results}

\begin{figure}
  \begin{center}
  \includegraphics[width=84mm]{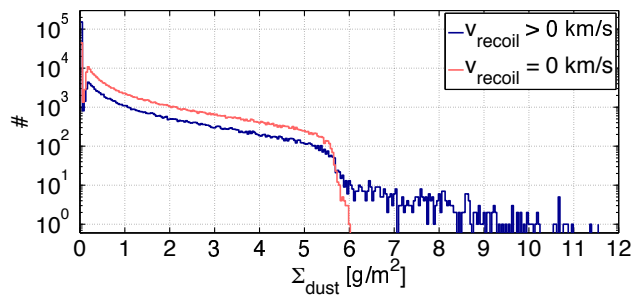}\\
   \caption{Histogram of dust column mass densities
     ($\Sigma_\mathrm{dust}$) for $2.5\times 10^{5}$ random
     observations of QSOs, obtained by integrating the mass density of
     the randomly oriented dust tori along the line of sight to the
     recoiling, off-centre SMBHs (dark blue curve). For reference, we
     show the same histogram for a similar sample of QSO without any
     recoil ($v_\mathrm{recoil}=0$; light red curve).  The recoiling
     QSOs are much more likely to be seen completely unobscured (there
     are $\sim 152700\ (\simeq 61$ per cent) and $\sim 44200\ (\simeq
     18$ per cent) QSOs with $\Sigma_\mathrm{dust}=0$ in the samples with and
     without recoil, respectively). The recoiling QSOs also show a
     tail of high $\Sigma_\mathrm{dust}$ values up to twice the
     maximum value for non-recoiling QSOs, as expected (see text).}
  \label{Fig:Sigma_Hists}
 \end{center}
 \end{figure}

We are now ready to present the results of the Monte Carlo simulation
of $2.5\times 10^{5}$ random observations of recoiling QSOs. In
Fig.~\ref{Fig:Sigma_Hists}, we show the histogram of dust column
mass densities ($\Sigma_\mathrm{dust}$) obtained by integrating the
mass density of the randomly oriented dust tori along the
line-of-sight to the recoiling SMBH (dark blue curve). For reference,
we also show a $\Sigma_\mathrm{dust}$ histogram for a second set of
simulations of $2.5\times 10^{5}$ QSOs with recoil velocities set to
$v_\mathrm{recoil}=0$ (light red curve). We have found that the total
number of unobscured QSOs (i.e. with $\Sigma_\mathrm{dust}=0$) is
$\sim 152,700\ (\simeq 61$ per cent) and $\sim 44,200\ (\simeq 18$ per cent) for the
$v_\mathrm{recoil}>0$ and $v_\mathrm{recoil}=0$ samples,
respectively. QSOs in the $v_\mathrm{recoil}=0$ sample do not have
$\Sigma_\mathrm{dust}$ values above $\Sigma_\mathrm{dust}\simeq
6\ \mathrm{g/m^2}$ because such QSOs are never obscured by more than a
half cross-section of their dust torus, and a half cross-section of
any dust tori in the \citet{Schartmannetal2005} model have a maximum
possible dust column density of $\Sigma_\mathrm{torus}\simeq
6\ \mathrm{g/m^2}$. Recoiling QSOs
($v_\mathrm{recoil}>0\ \mathrm{km/s}$) can obtain up to twice higher
values of $\Sigma_\mathrm{dust}$ (i.e. up to
$\Sigma_\mathrm{dust}\simeq 12\ \mathrm{g/m^2}$), which corresponds to
the maximal obscuration of a QSO located in the `equatorial plane',
behind the dust torus (and for the geometrically largest tori). 
This configuration is rare, with only $409\ (\simeq 0.2$ per cent) QSOs in
the $v_\mathrm{recoil}>0$ sample having $\Sigma_\mathrm{dust}>
6\ \mathrm{g/m^2}$.

\begin{figure}
  \begin{center}
  \includegraphics[width=84mm]{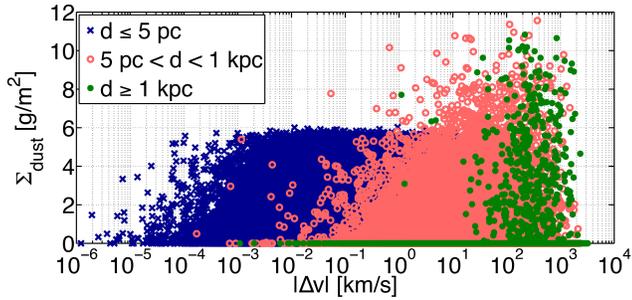}\\
   \caption{The line-of-sight dust column density,
     $\Sigma_\mathrm{dust}$, versus the (absolute value) of
     line-of-sight velocity offset $|\Delta v|$ relative to the host
     galaxy, for each of the $2.5\times 10^{5}$ recoiling QSO.  For
     completeness, we show the entire range of velocities we
     simulated, down to unmeasurably low values.  To illustrate how
     far each recoiling QSO has travelled from the galactic centre by
     the time of the mock observation, we sorted the data into three
     groups, based on their radial offset $d$ from the galactic
     centre: QSOs with the smallest offset ($d\leq 5\ \mathrm{pc}$;
     blue crosses), QSOs with offsets within roughly an order of
     magnitude of the size of their parent torus ($5\ \mathrm{pc} < d
     < 1\ \mathrm{kpc}$; red circles), and the most distant QSOs
     ($d\geq 1\ \mathrm{kpc}$; green full circles). The distribution
     of points shows three distinct $\Sigma_\mathrm{dust}$ values:
     $\Sigma_\mathrm{dust}=0$ (corresponding to the ejected QSOs with
     the largest offsets, viewed in an unobscured direction),
     $\Sigma_\mathrm{dust} \simeq 6\ \mathrm{g/m^2}$
     (the maximum possible obscuration in the \citet{Schartmannetal2005} model by half of the torus,
     for QSOs whose orbits have fully decayed back to
     the centre by dynamical friction and are viewed along the
     equator), and at
     $\Sigma_\mathrm{dust} \simeq 12\ \mathrm{g/m^2}$
     (the maximum possible obscuration, for QSOs located behind the
     full torus).}
  \label{Fig:NH_vs_vr}
 \end{center}
 \end{figure}

Fig.~\ref{Fig:NH_vs_vr} shows $\Sigma_\mathrm{dust}$ versus
$|\Delta v|$ for the $2.5\times 10^5$ recoiling QSOs. We have
calculated the sample correlation coefficient, $r$, between
$\Sigma_\mathrm{dust}$ and $\log_{10}(|\Delta v|)$, defined by 
\begin{equation}\label{eq:rcoeff}
r \equiv \frac{ \langle \Sigma_\mathrm{dust} \log_{10}(|\Delta v|)\rangle -\langle \Sigma_\mathrm{dust} \rangle \langle \log_{10}(|\Delta v|) \rangle}{\sqrt{\langle \Sigma_\mathrm{dust}^2\rangle-\langle \Sigma_\mathrm{dust} \rangle^2} \sqrt{\langle \log_{10}(|\Delta v|)^2\rangle-\langle \log_{10}(|\Delta v|) \rangle^2}},
\end{equation}
restricted to various ranges of $|\Delta v|_\mathrm{min} \leq |\Delta
v| \leq |\Delta v|_\mathrm{max}$. Here $\langle...\rangle$ refers to
averaging over the sample of $2.5\times 10^5$ QSOs, or its various
subsets.

\begin{figure}
  \begin{center}
  \includegraphics[width=84mm]{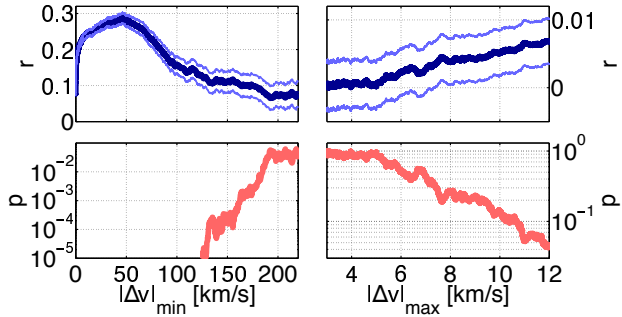}\\
   \caption{Sample correlation coefficient $r$ between
     $\Sigma_\mathrm{dust}$ and $\log_{10}(|\Delta v|)$, with its $1\sigma$
     uncertainty (upper panels, blue curves), and the corresponding
     $p$ values for the hypothesis of no correlation (lower panels,
     red curves).  The left-hand (right-hand) panels show both $r$ and $p$ as a
     function of $|\Delta v|_\mathrm{min}$ ($|\Delta
     v|_\mathrm{max}$), for sub-samples of obscured QSOs with $|\Delta
     v|\geq |\Delta v|_\mathrm{min}$ ($|\Delta v|< |\Delta
     v|_\mathrm{max}$). QSOs with $|\Delta v|\gsim 5\ \mathrm{km/s}$
     show a positive correlation, with $r$ as high as $r=0.28\pm 0.02$
     for $|\Delta v|_\mathrm{min}\approx
     45\ \mathrm{km/s}$.}
  \label{Fig:r_p_vs_vcut}
 \end{center}
 \end{figure}

Fig.~\ref{Fig:r_p_vs_vcut} shows $r$ as a function of $|\Delta
v|_\mathrm{min}$ (left-hand panels) and $|\Delta v|_\mathrm{max}$ (right-hand
panels) for sub-samples of obscured QSOs with $|\Delta v|\geq |\Delta
v|_\mathrm{min}$ and $|\Delta v|< |\Delta v|_\mathrm{max}$,
respectively. As can be seen in the figure, obscured QSOs with
$|\Delta v|< 5\ \mathrm{km/s}$ show no correlation between their
$\Sigma_\mathrm{dust}$ and $|\Delta v|$ values ($r_{<5}\simeq 0$, and the corresponding
$p$--value for the hypothesis of no correlation is $p\simeq 0.95$), while QSOs with
$|\Delta v|\geq 5\ \mathrm{km/s}$ show a significant correlation, with $r_{5} = 0.22\pm 0.01$ and $p\simeq 0$.  
The correlation coefficient increases with $|\Delta v|_\mathrm{min}$
until it reaches its maximum at $|\Delta
v|_\mathrm{min}=45\ \mathrm{km/s}$ with $r_{45} = 0.28\pm 0.02$.

Allowing $|\Delta v|_\mathrm{min}$ and $|\Delta v|_\mathrm{max}$ to
vary simultaneously does not change the result that $r$ is the highest
for the $3,824$ obscured QSOs with $|\Delta v|\geq 45\ \mathrm{km/s}$.
Depending on the signal to noise, the measurement error on $|\Delta
v|$ of individual QSOs is a few tens of km/s (for example from
spectral line fitting, or by using the centroids of individual broad
lines; see e.g., \citealt{Juetal2013,Shenetal2013}). We therefore
restrict our further investigations to the $3824$ obscured QSOs with
$|\Delta v|\geq 45\ \mathrm{km/s}$. For this subset, we have found the
approximate relation $\Sigma_\mathrm{dust}\left(\log_{10}\{|\Delta
v|\} \right)=\left (1.7\pm 0.1\right )\ \mathrm{g/m^2}\times \left
(\log_{10}\{|\Delta v|/[1\ \mathrm{km/s}]\}\right ) - \left (1.3\pm
0.2\right )\ \mathrm{g/m^2}$, although of course this relation depends
strongly on the underlying set of models. We also note that in
addition to the random measurement errors on $|\Delta v|$, broad line
centroids can shift by $\cal{O}$(100 km/s) between pairs of
observations only weeks apart, for physical reasons unrelated to
recoil \citep{Juetal2013,Shenetal2013}.  These shifts, as well as the
measurement errors, are not expected to correlate with
$\Sigma_\mathrm{dust}$. However, they introduce random errors on the
$\Sigma_\mathrm{dust}\left (\log_{10}(|\Delta v|) \right )$ relation,
and can therefore increase the number of QSOs required to detect the
underlying $\Sigma_\mathrm{dust}$--$\log_{10}(|\Delta v|)$ correlation
(see discussion below).

\begin{figure}
  \begin{center}
  \includegraphics[width=84mm]{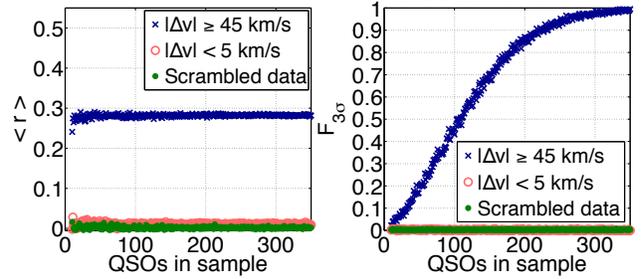}\\
   \caption{Left-hand panel: the expectation value $\langle r\rangle$
     of the correlation coefficient between $\Sigma_\mathrm{dust}$ and
     $\log_{10}(|\Delta v|)$, computed in 1000 random samples of $N$ QSOs.  The
     results are shown as a function of $N$ in the range $N\in
     [10,350]$ for three different QSO data sets: obscured QSOs with
     $|\Delta v|\geq 45\ \mathrm{km/s}$ (blue crosses), obscured QSOs
     with $|\Delta v|< 5\ \mathrm{km/s}$ (red circles), and QSOs
     without recoil, but with $|\Delta v|$ values assigned randomly
     from the $|\Delta v|$ distribution for recoiling SMBHs
     (`scrambled data'; filled green circles). While $\langle
     r\rangle\simeq 0$ for all $N$ in the last two data sets, the
     first data set shows a stable correlation with $\langle
     r\rangle\simeq 0.28$ for all $N$.
     Right-hand panel: the fraction $F_{3\sigma}$ of the 1,000 random
     samples in which the `no correlation' null hypothesis can be
     rejected at $\geq 3\sigma$ significance.  While
     $F_{3\sigma}\simeq 0$ for the $|\Delta v|< 5\ \mathrm{km/s}$ and
     `scrambled' data sets, it increases monotonically with $N$ from
     $F_{3\sigma}\simeq 0$ to $1$ for the $|\Delta v|\geq
     45\ \mathrm{km/s}$ data set, rising above $F_{3\sigma}\geq 0.95$
     for $N\geq 260$.}
  \label{Fig:CorrCoeff}
 \end{center}
 \end{figure}

We next estimate the minimum number of QSOs, $N_\mathrm{min}$,
required for a significant detection of the correlation. To do this,
for each trial $N\in [10,350]$ QSOs, we (i) generated $1000$
independent random samples of $N$ QSOs, selected from among the
$3824$ obscured QSOs recoiling with $|\Delta v|\geq
45\ \mathrm{km/s}$, (ii) computed the $p$-value of the hypothesis of
no correlation between $\Sigma_\mathrm{dust}$ and $|\Delta v|$ for
each of these 1000 samples, and (iii) measured the fraction of
samples for which $p<3\times 10^{-3}$ (corresponding to the rejection
of the no-correlation hypothesis at $3\sigma$ confidence). This
fraction, $F_{3\sigma}$, is shown in Fig.~\ref{Fig:CorrCoeff} (blue
crosses, right-hand panel), together with the average $\langle r\rangle$
over the $1000$ random samples, as a function of $N$ (blue crosses,
left-hand panel).

As a test of these statistics, we have repeated the calculations of
both $\langle r\rangle$ and $F_{3\sigma}$ for two additional sequences
of 1000 random samples of $N\in [10,350]$ QSOs.  In the first set, we
began with recoiling QSOs with $|\Delta v|<5\ \mathrm{km/s}$; the
results are shown by the red empty circles in
Fig.~\ref{Fig:CorrCoeff}).  In the second set, we computed random
$\Sigma_\mathrm{dust}$ values for stationary SMBHs ($v_\mathrm{recoil}
= 0$), and associated a $|\Delta v|$ with each case, drawn randomly
from the $|\Delta v|$ distribution for the recoiling
($v_\mathrm{recoil} > 0$) SMBHs. The results for these `scrambled'
data'' are shown in Fig.~\ref{Fig:CorrCoeff} by the filled green
circles.

As can be seen in Fig.~\ref{Fig:CorrCoeff}, neither the $|\Delta v|<
5\ \mathrm{km/s}$, nor the `scrambled data' sets show correlations,
i.e. $\langle r\rangle\simeq 0$ and $F_{3\sigma}\simeq 0$ for all $N$.
On the other hand, for obscured QSOs with $|\Delta v|\geq
45\ \mathrm{km/s}$, we find $\langle r\rangle\simeq 0.28$ for all
$N$s, and $F_{3\sigma}$ increases monotonically with $N$ from
$F_{3\sigma}\simeq 0$ to $F_{3\sigma}\simeq 1$ within the interval
$N\in[10,350]$. We conclude that the hypothesis of no correlation
between $\Sigma_\mathrm{dust}$ and $\log_{10}(|\Delta v|)$ could be rejected with
a $3\sigma$ significance $95$ per cent of the time by observing a random
sample of $\sim 260$ obscured QSOs with $|\Delta v|\geq
45\ \mathrm{km/s}$. Note that in our simulations, $3824$ QSOs
($\simeq 1.5$ per cent) were obscured having $|\Delta v|\geq
45\ \mathrm{km/s}$, thus gathering a sample of $\sim 260$ such QSOs
may require observing as many as $\sim 17000$ recoiling QSOs (and a
larger number of QSOs in general, accounting for the fact that not all
QSOs experienced a recoil in the past).
 
\begin{figure}
  \begin{center}
  \includegraphics[width=84mm]{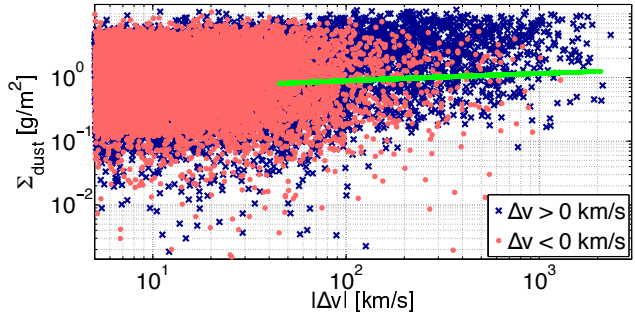}\\
   \caption{
    $\Sigma_\mathrm{dust}$ versus $|\Delta v|$ for the
     $17157$ obscured QSOs with $|\Delta v|\geq 5\ \mathrm{km/s}$.
     Receding (i.e. $\Delta v>0$) QSOs are shown by blue crosses,
     while approaching QSOs (i.e. $\Delta v<0$) are shown by red filled
     circles. We also show the $\Sigma_\mathrm{dust}\left (\log_{10}(|\Delta v|)
\right )=\left (1.7\pm 0.1\right )\ \mathrm{g/m^2}\times \left (\log_{10}(|\Delta v|/[1\ \mathrm{km/s}])\right ) 
- \left (1.3\pm 0.2\right )\ \mathrm{g/m^2}$ curve (green solid line) we obtained by fitting $\Sigma_\mathrm{dust}$ versus 
     $\log_{10}(|\Delta v|)$ values for obscured QSOs with $|\Delta v|\geq 45\ \mathrm{km/s}$.
     Receding SMBHs are slightly more common among the
     obscured ($\Sigma_\mathrm{dust}>0$) QSOs, their velocity offsets
     extend to higher values, and they tend to be more heavily
     obscured (larger $\Sigma_\mathrm{dust}$), especially at large
     velocity offsets ($\Delta v\gsim 100\ \mathrm{km/s}$). See text
     for the origin of these differences.}
  \label{Fig:NH_vs_vr_posneg}
 \end{center}
 \end{figure}

To examine the impact of the sign of the line-of-sight velocity
offset, in Fig.~\ref{Fig:NH_vs_vr_posneg} we show
$\Sigma_\mathrm{dust}$ versus $|\Delta v|$ in the subsample of
$17157$ obscured QSOs with $|\Delta v|\geq 5\ \mathrm{km/s}$, and
indicate whether they are moving away from ($\Delta v>0$, shown as
blue crosses) or towards ($\Delta v<0$, shown as red full circles) the
observer. First, we note that receding QSOs are slightly more
common, with an excess of $1405$ and $1162$ in the $|\Delta v|\geq
5\ \mathrm{km/s}$, and $|\Delta v|\geq 45\ \mathrm{km/s}$ samples,
respectively, over the approaching cases.
As Fig.~\ref{Fig:NH_vs_vr_posneg} shows, $\Sigma_\mathrm{dust}$
values of the receding QSOs also tend to be higher, especially when
the velocity offset is large ($|\Delta v|\gsim 100\ \mathrm{km/s}$).
This is because receding QSOs are more likely to be found behind,
rather than in front of the tori. In order for a SMBH to be
approaching with a large line-of-sight velocity while being located
behind the torus, it would have had to reach its first U-turn. As
mentioned above, the majority of SMBHs with large recoil velocities do
not make their first U-turn within a QSO lifetime (or within our mock
observation times).
 
\begin{figure}
  \begin{center}
  \includegraphics[width=84mm]{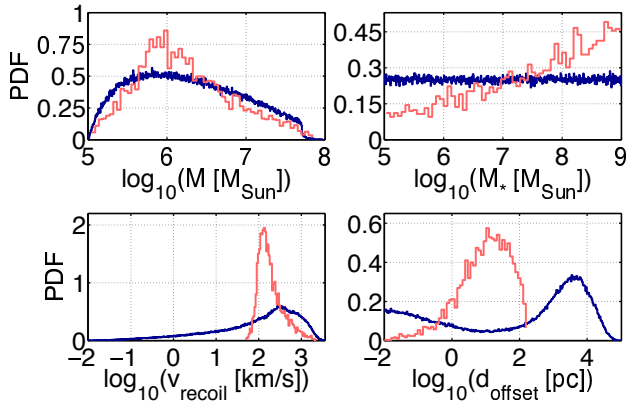}\\
   \caption{The probability distribution functions of different
     parameters of the QSOs within the entire sample (blue curves),
     and within the subsample of obscured QSOs with significant
     correlations (i.e. with $|\Delta v|\geq 45\ \mathrm{km/s}$; red
     curves). The four different panels, clockwise from top left, show
     the SMBH masses ($M$), the masses of the nuclear star clusters
     ($M_{*}$), recoil velocities ($v_\mathrm{recoil}$), and the
     offsets of QSOs from the centres of their host galaxies
     ($d_\mathrm{offset}$). The difference between the red and blue
     curves reveal that QSOs with $M\simeq(0.5--3)\times 10^6
     M_{\odot}$, $M_{*}\gtrsim 10^7 M_{\odot}$,
     $v_\mathrm{recoil}\simeq 100\ \mathrm{km/s}$, and
     $d_\mathrm{offset} \simeq 10^{-1}-10^{2}\ \mathrm{pc}$ are the
     most over-represented in the sample showing strong correlations
     between $\Sigma_\mathrm{dust}$ and $|\Delta v|$.}
  \label{Fig:Ratio}
 \end{center}
 \end{figure}

In Fig.~\ref{Fig:Ratio}, we compare the numerically constructed probability
density functions of various parameter distributions of QSOs within the whole
sample, and within the subsample of obscured QSOs with $|\Delta v|\geq 45\ \mathrm{km/s}$.
Based on the comparison, we conclude that AGNs where the central SMBH
has a mass around $M\simeq(0.5--3)\times 10^6 M_{\odot}$, a positional offset
from the centre of the host galaxy between $d_\mathrm{offset} \simeq 10^{-1}-10^{2}\ \mathrm{pc}$,
and having a nuclear star cluster with mass $M_{*}\gtrsim 10^7 M_{\odot}$,
have the highest probability of being in the subset of obscured QSOs with
$|\Delta v|\geq 45\ \mathrm{km/s}$. The fact that high-mass nuclear star
clusters are over-represented in the subset of QSOs showing $\Sigma_\mathrm{dust}$--$|\Delta v|$
correlation (see the upper-right panel in Fig.~\ref{Fig:Ratio}) suggests that
such QSOs are more likely to be found in active galaxies with higher masses.
We attribute this to the fact that in more massive galaxies, the recoiling BHs
spend longer times at off-centre distances comparable to the size of the obscuring
torus. This host-size dependence of the proposed correlation is another
potential testable prediction of our toy model.
 
\begin{figure}
  \begin{center}
  \includegraphics[width=84mm]{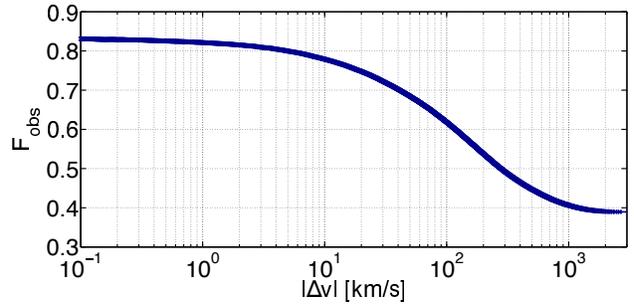}\\
   \caption{The fraction of obscured ($\Sigma_\mathrm{dust}>0$)
     sources among the entire sample of $2.5\times10^5$ QSOs. The
     trend of $\cal{F_{\rm obs}}$$\left (|\Delta v|\right )$
     decreasing with $|\Delta v|$ from $\cal{F_{\rm obs}}$$\left
     (<10\ \mathrm{km/s}\right )\gtrsim 0.8$ to $\cal{F_{\rm
         obs}}$$\left (>10^3\ \mathrm{km/s}\right )\lesssim 0.4$ could
     provide an independent test for a chosen combination of recoil,
     trajectory, and dust tori models, when compared with the observed
     fraction of obscured (e.g. type~II) QSOs in the future.}
  \label{Fig:R}
 \end{center}
 \end{figure}

Finally, we have studied the fraction of obscured QSOs within the
whole sample of QSOs, $\cal{F_{\rm obs}}$$\left (|\Delta v|\right )$.
We have found that this fraction decreases monotonically with $|\Delta
v|$ from $\cal{F_{\rm obs}}$$\left (<10\ \mathrm{km/s}\right )\gtrsim
0.8$ to $\cal{F_{\rm obs}}$$\left (>10^3\ \mathrm{km/s}\right
)\lesssim 0.4$ (see Fig.~\ref{Fig:R}). This result can be compared
in the future to the observed fraction of obscured (e.g. type~II)
QSOs, which can potentially provide an independent test for a chosen
combination of recoil, trajectory, and dust tori models.
   
\section{Discussion}
\label{sec:Discussion}

The results in the previous section show that a positive correlation may
be present between $\Sigma_\mathrm{dust}$ and $|\Delta v|$.  We
emphasize that this conclusion is based on a specific set of idealized
toy models -- more realistic models will undoubtedly lead to different
predictions.  Nevertheless, our results should motivate an
observational search; more realistic and flexible models will likely
be needed to interpret any correlation discovered in actual data.  We
plan to present the results from a search in the SDSS DR10 data base in
a forthcoming publication.

Here we address two natural caveats: while proxies can be used to
estimate both $\Sigma_\mathrm{dust}$ and $|\Delta v|$, these estimates
will have uncertainties we have so far ignored. Although we have no
a priori reason to expect that these uncertainties will be correlated,
they will represent noise, which will reduce the statistical
significance of any correlation, and thus increase the number of
QSOs required for a detection. Furthermore, we have so far assumed
that a clean QSO sub-sample with $\Sigma_\mathrm{dust}>0$ and $|\Delta
v|\geq 45\ \mathrm{km/s}$ can be constructed. More realistically,
observational errors will pollute any such selection (e.g. QSOs with
$|\Delta v|\geq 45\ \mathrm{km/s}$ can be missed, and QSOs with
$|\Delta v|< 45\ \mathrm{km/s}$ can be mistakenly included).

The line-of-sight velocity offset ($|\Delta v|$) can be estimated using
the centroids or peaks of broad lines, relative to those of the narrow
lines. The level of obscuration ($\Sigma_\mathrm{dust}$) can be estimated
using QSO colours: a larger obscuring column should result in a
redder continuum (see e.g. \citet{Ledouxetal2015} and references therein). An alternative proxy for the obscuring column
could be the equivalent width of the broad lines themselves.
Although the geometry of the broad line region is not fully
understood, type II quasars have weak or no broad emission lines. In
the standard unified model of AGN \citep{Antonucci1993}, this is attributed to the
obscuration of the high-velocity broad-line region by the dusty torus
(note that the continuum is still visible and can therefore not be
spatially colocated with the broad-line emitting region). This
implies that the equivalent width of these broad lines must vary with
the level of the obscuration.

Here we estimate the impact of having random errors on both
$|\Delta v|$ and $\Sigma_\mathrm{dust}$. 
As mentioned above, we expect $|\Delta v|$, when
determined using line centroids, to suffer physical variations of
order $\sigma_{\Delta v}\approx 100\ \mathrm{km/s}$.  We therefore
repeated our analysis above, except we added a random additional
$\Delta v$ component to the velocity offset of each simulated QSO,
drawn from a Gaussian distribution with a standard deviation of
$\sigma_{\Delta v}$.  For any given $\sigma_{\Delta v}$, we then
selected the subset of the $2.5\times10^5$ QSOs with
$\Sigma_\mathrm{dust}>0$ and $|\Delta v|\geq 45\ \mathrm{km/s}$, as
before. Finally, we measured the expectation value of the correlation
coefficient $\langle r \rangle$ between $\Sigma_\mathrm{dust}>0$ and
$|\Delta v|$ in 1000 randomly drawn sets of $N$ QSOs from this
subset, and determined the minimum number $N_{\rm min}$ of QSOs
required to infer $r>0$ with $3\sigma$ significance $95$ per cent of the
time.

\begin{figure}
  \begin{center}
  \includegraphics[width=84mm]{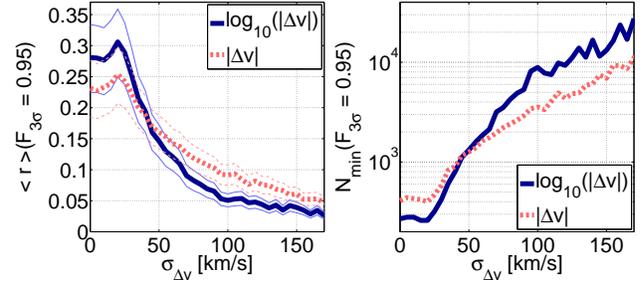}\\
   \caption{The figure shows the impact of random errors in the
     velocity offset; the error is assumed to be Gaussian with a
     standard deviation of $\sigma_{\Delta v}$. The right-hand panel shows
     the increase in the minimum number of QSOs ($N_{\rm min}$)
     required to detect correlations between $\Sigma_\mathrm{dust}$
     and $\log_{10}(|\Delta v|)$ (blue solid curves), or between
     $\Sigma_\mathrm{dust}$ and $|\Delta v|$ (red dashed curves) in a
     subset of QSOs with $\Sigma_\mathrm{dust}>0$ and $|\Delta
     v|\geq 45\ \mathrm{km/s}$, as $\sigma_{\Delta v}$
     increases. Curves in the left-hand panel, with $\pm1\sigma$ errors,
     show the corresponding decrease in the correlation coefficient
     $\langle r \rangle$ (see text for more details). Note that
     measurement errors below $\sigma_{\Delta v}\leq
     45\ \mathrm{km/s}$ results with higher $\langle r \rangle$ and
     lower $N_{\rm min}$ values for the
     $\Sigma_\mathrm{dust}$--$\log_{10}(|\Delta v|)$ correlation,
     while the $\Sigma_\mathrm{dust}$--$|\Delta v|$ correlation
     appears to be stronger when $\sigma_{\Delta v}\geq
     45\ \mathrm{km/s}$.
}
  \label{Fig:v-error}
 \end{center}
 \end{figure}

The result of this random-velocity-error exercise is shown in
Fig.~\ref{Fig:v-error} for both
$\Sigma_\mathrm{dust}$--$\log_{10}(|\Delta v|)$ and
$\Sigma_\mathrm{dust}$--$|\Delta v|$ correlations. The right-hand panel in
this figure shows $N_{\rm min}$ as a function of the random error
$\sigma_{\Delta v}$. As expected, the number of QSOs required to
detect a correlation increase monotonically with $\sigma_{\Delta v}$,
rising to $N_{\rm min}=3560$ for the $\Sigma_\mathrm{dust}$--$|\Delta
v|$ correlation and for $\sigma_{\Delta v}= 100\ \mathrm{km/s}$; this
is more than an order of magnitude increase over the $N_{\rm min}=260$ in the
idealized case for the $\Sigma_\mathrm{dust}$--$\log_{10}(|\Delta v|)$
correlation without any velocity-offset error. The left-hand panel in
Fig.~\ref{Fig:v-error} shows the corresponding values of the
correlation coefficients; $\langle r \rangle$ decreases monotonically
with $\sigma_{\Delta v}$, reaching $\langle r\rangle= 0.091\pm 0.016$
for the $\Sigma_\mathrm{dust}$--$|\Delta v|$ correlation and for
$\sigma_{\Delta v}= 100\ \mathrm{km/s}$. This is a decrease from the
$\langle r\rangle= 0.28\pm 0.02$ in the idealized case for the
$\Sigma_\mathrm{dust}$--$\log_{10}(|\Delta v|)$ correlation by a
factor of three. 
Additionally, we have calculated $\langle r \rangle$ for partially obscured
QSOs only (i.e. QSOs with $0<\Sigma_\mathrm{dust}\lsim
2.3\ \mathrm{g/m^2}$), again assuming $\sigma_{\Delta v}= 100\ \mathrm{km/s}$, and we have
found it to be $\langle r \rangle = 0.026\pm 0.004$, measured between $|\Delta v|$ and
$\Sigma_\mathrm{dust}$.

Similarly to the errors in $|\Delta v|$, we have studied the impact of
errors in $\Sigma_\mathrm{dust}$. Here, there are several potential
concerns. First, in the presence of uncertainties, a sharp selection
$\Sigma_\mathrm{dust}>0$ is unrealistic. Secondly, for large obscuring
columns, with $\Sigma_\mathrm{dust}$ comparable to the maximum possible column
density of a half-torus according to the \citet{Schartmannetal2005} model
(see Fig.~\ref{Fig:NH_vs_vr}), $\Sigma_\mathrm{torus}\simeq 6\ \mathrm{g/m^2}$ (i.e. for
configurations corresponding to true type II QSOs), the broad lines
may be undetectable, or may have too low S/N for a reliable centroid
measurement.\footnote{Note that the broad lines and the optical
  continuum are thought to arise from different spatial locations, and
  can suffer different levels of obscuration; we ignore this
  complication here.} We here repeated our fiducial analysis, except
that we replaced the selection criterion $\Sigma_\mathrm{dust}>0$ by
the range $\Sigma_{\rm min} < \Sigma_{\rm dust} < \Sigma_{\rm max}$.
We chose $\Sigma_{\rm min}=0.6\ \mathrm{g/m^2}\approx
0.1\times\Sigma_\mathrm{torus}$ and $\Sigma_{\rm
  max}=2.3\ \mathrm{g/m^2}\approx
0.4\times\Sigma_\mathrm{torus}$. The number of QSOs within this
range (and also satisfying $\Delta v \geq 45$ km/s as before) is $N =
1436$, or $\sim0.6$ per cent of the full sample of $2.5\times10^5$ QSOs. With this
cut, we find that the correlations remain significant: $\langle
r\rangle = 0.098\pm 0.026$ for the $\Sigma_\mathrm{dust}$--$\log_{10}(|\Delta v|)$ correlation, with a $p$-value of $p=2\times10^{-4}$.

We also evaluated the impact of random Gaussian errors in
$\Sigma_{\rm dust}>0$ with a standard deviation of $\sigma_\Sigma$. The
results are similar to those shown in Fig.~\ref{Fig:v-error}: we
find that $\langle r \rangle$ decreases, and the minimum number of
QSOs required to detect the correlation between $\Sigma_{\rm dust}$ and $\log_{10}(|\Delta v|)$ ($N_\mathrm{min}$) increases with
$\sigma_\Sigma$. For the case of $\sigma_\Sigma =
0.6\ \mathrm{g/m^2}= 0.1\times\Sigma_\mathrm{torus}$, we
find $\langle r\rangle = 0.27\pm 0.05$, with a $p$-value of $p\approx 10^{-58}$,
and $N_\mathrm{min}=290$. Thus, random Gaussian errors in
$\Sigma_{\rm dust}>0$ have a very small effect on the $\Sigma_{\rm dust}$--$\log_{10}(|\Delta v|)$ correlation.

Based on the above, we conclude that statistical errors on estimating
the velocity offsets and obscuring column densities can diminish the
correlations predicted in the idealized toy model. This is mostly due
to the impact of these errors on the purity of selection of a
sub-sample, intended to include only the most correlated QSOs. In the
examples above, we find that this increases the number of QSOs
required for a detection by an order of magnitude, but there is no
indication that these errors will prevent a detection of the
correlations in a large sample (tens of thousands) of QSOs.

Finally, as the distribution of SMBH masses in QSOs observable in
different survey projects may vary significantly, we have considered
how using higher SMBH masses (see \citealt{GreeneHo2007,GreeneHo2009} and \citealt{ShenKelly2012}) affect our results.
In these tests, we have also taken into account the possibility of a correlation
between the SMBH mass and the mass of the stellar core \citep{Ferrareseetal2006,Grahametal2011,KormendyHo2013}.
We have found that the increased SMBH masses lead fewer SMBHs
escaping the galactic nucleus (an effect that should increase the number of
SMBHs in the population showing $\Sigma_\mathrm{dust}$--$|\Delta v|$ correlation), however
they also lead to significantly smaller torus sizes, resulting with a net decrease in
the correlation coefficient of a factor of $\sim 2$. This demonstrates that although the
strength of the $\Sigma_\mathrm{dust}$--$|\Delta v|$ is sensitive to the underlying assumptions
of the applied models, the $\Sigma_\mathrm{dust}$--$|\Delta v|$ correlation itself is a robust result.

\section{Conclusions}
\label{sec:Conclusions}

A large fraction of luminous QSOs are believed to be triggered by
merger events, and if QSO activity often follows the coalescence of
SMBHs in the merging nuclei, then a large fraction of all QSOs may
have experienced recent gravitational recoil. While a handful of
QSOs with large spatial or velocity offsets have been identified as
recoil candidates \citep{Komossa2012}, 
the ubiquity of this phenomenon remains poorly
known.

Here, we proposed a statistical technique to search for a population of
recoiling SMBHs with smaller velocity offsets among luminous QSOs,
based on a positive correlation between these velocity offsets and the
column densities of obscuring dust tori.  The correlations are
introduced by the damped oscillating motions of SMBHs with typical
recoil speeds, which are comparable to the escape velocities from
galactic nuclei. These SMBHs spend a significant time off-nucleus, and
at distances comparable to the size of obscuring tori in located in
the nuclei.

We have demonstrated, using simple models of gravitational
recoil, SMBH trajectories, and the geometry of obscuring dust tori,
that observing a random sample of a few thousand partially obscured
QSOs, with line-of-sight velocity offsets $|\Delta v|\geq
45\ \mathrm{km/s}$, could allow a significant detection of the
correlation between the line of sight dust column density
$\Sigma_\mathrm{dust}$ and $|\Delta v|$ (or $\log_{10}(|\Delta v|)$).

The existence of this correlation is found to be robust within the
simple models we choose, and detectable even in the face of random
errors on $\Sigma_\mathrm{dust}$ and $|\Delta v|$. However, we expect
that the tightness of the correlation, the underlying
$\Sigma_\mathrm{dust}-|\Delta v|$ relation, and the properties of the
QSO subset showing correlations can depend strongly on the model
assumptions. Nevertheless, our results should motivate searching for
correlations in real data; a positive detection would allow testing
various combinations of recoil, BH trajectory, and dust tori
models. A further test could be possible by comparing the predicted
fraction of obscured QSOs to the observed fraction of type II QSOs.
We propose to carry out these tests in the future using catalogues of
observational data for QSOs, such as the SDSS-DR10 QSO catalogue.

\section*{Acknowledgements}

The authors would like to thank Bence B\'ecsy, Gergely D\'alya, and
\'Akos Sz\"olgy\'en for useful discussions and for valuable comments on
the manuscript. P\'eter Raffai is grateful for the support of the Hungarian
Academy of Sciences through the 'Bolyai J\'anos' Research Scholarship
programme. We also acknowledge financial support from the NASA 
Astrophysics Theory Program under grant no. NNX11AE05G (to ZH) and
from OTKA under grant no. 101666 (to ZF).

\label{lastpage}

\end{document}